# A Task Based Sensor-Centric Model for Overall Energy Consumption

Najmeh Kamyabpour
INEXT Centre for Innovation in IT Services and Applications
University of Technology, Sydney
najmeh@it.uts.edu.au

Doan B. Hoang
INEXT Centre for Innovation in IT Services and Applications
University of Technology, Sydney
dhoang@it.uts.edu.au

*Abstract*— **Sensors have limited resources so it is important to manage their resources efficiently to maximize their use. A sensor's battery is a crucial resource as it singly determines the lifetime of sensor network applications. Since these devices are useful only when they are able to communicate with the world, radio transceiver of a sensor as an I/O and a costly unit plays a key role in its lifetime. This resource often consumes a big portion of the sensor's energy as it must be active most of the time to announce the existence of the sensor in the network. As such the radio component has to deal with its embedded sensor network whose parameters and operations have significant effects on the sensor's lifetime. In existing energy models, hardware is considered, but the environment and the network's parameters did not receive adequate attention. Energy consumption components of traditional network architecture are often considered individually and separately, and their influences on each other have not been considered in these approaches. In this paper we consider all possible tasks of a sensor in its embedded network and propose an energy management model. We categorize these tasks in five energy consuming constituents. The sensor's Energy Consumption (EC) is modeled on its energy consuming constituents and their input parameters and tasks. The sensor's EC can thus be reduced by managing and executing efficiently the tasks of its constituents. The proposed approach can be effective for power management, and it also can be used to guide the design of energy efficient wireless sensor networks through network parameterization and optimization.**

## I. INTRODUCTION

The importance of the power management for sensors is well known, and many specific network protocols are attempted to reduce the energy consumption of wireless sensor networks. It would be simple to optimize the total energy consumption of a sensor network application if one can attribute the energy consumption to a particular component of a sensor associated with a particular activity and the role of the sensor within the application. However, that has been proven difficult. Existing power management approaches mainly optimize the Energy Consumption (EC) strictly along the OSI layers in isolation hence they are not able to minimize the overall energy consumption of a sensor network application. The EC in one network protocol layer cannot be separated from the overall EC. In fact, minimization of the EC in one network layer may increase the EC of other network layers. For example, turning off and on a sensor as an energy minimization technique in the physical layer creates the necessity of scheduling. Moreover, the clustering procedure at the network layer causes excessive exchange of messages during the clustering process and hence dissipates a considerable amount of energy for message transmission.

Efforts in minimizing the EC have increased over the last few years, however, they mostly focused on some specific and separated components of energy dissipation based on the layer architecture such as MAC protocols[1],[2], routing[3], topology management[4] and data aggregation[5]. Minimizing the EC of one layer may increase the energy requirements of other layers and hence may not guarantee the minimization of the overall EC of the entire network. The cross layer idea aims to enhance the performance of the system by jointly optimizing multiple protocol layers[6]. It is argued that Cross-Layer Designs with tight coupling between the layers become hard to review and redesign.

Our approach in this paper can be considered as a sensor-centric approach that takes into account a sensor's constituents and their energy-consuming activities (or tasks) in performing its role within the sensor network and the associated application. As a result, the architecture has a modular structure, yet embraces cross layer ideas. The proposed EC model is used for overall EC minimization and power management for sensor's resources. We will show how this model helps a sensor to manage power usage and lengthen its life in the network. We assume five energy consuming constituents: Individual, Local, Global, Environment, and Sink (figure 1). Starting from an individual sensor, the Individual constituent within the first circle represents all the activities the sensor has to do to survive and perform its sensing function. The Local constituent within the ring between the first and the second circle represents all the activities the sensor has to perform to build a relationship with its neighbor. The Global constituent within the ring between the second and the third circle represents all the activities the sensor has to perform to establish possible transport paths and carry data from itself to the destination (sink). The Sink constituent within the thirds and the fourth circle represents all the activities the sensor has to perform as directed by the sink.

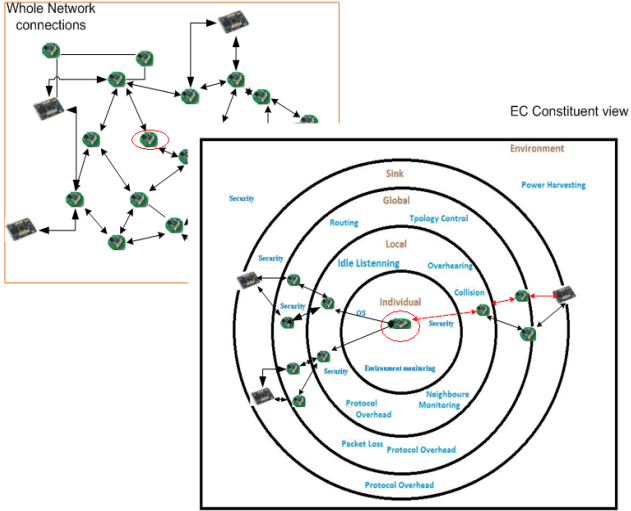

*Figure 1. Sensor centric view of a Wireless Sensor Network*

The final constituent, the Environment represents the activities the sensor may perform to harvest energy available from the environment. The EC minimization based on the constituents allows identifying sensor workload attributed to constituent, improving resource utilization through selection and load balancing among constituents, and reducing power usage.

In principle, the constituent power can be metered by tracking each hardware resources used by a constituent task and converting the resource usage to the power usage based on a power model for the resource. Our approach does not require any additional instrumentation of the application workload or operating system within the constituents. The constituent basis approach can naturally adapt to changes in applications and even hardware configurations. While prior works have proposed mechanisms to design energy-efficient individual network protocol or network layer, they are not capable of optimizing the overall EC of a sensor within the application. Generally a sensor has responsibility to process and execute assigned tasks while it has enough power. This constitutes a basis for our model that covers all possible energy consuming constituents. The sensor battery lifetime depends on how its functional tasks are distributed and executed among its Individual, Local, Global, Environment, and Sink constituents. To execute a task, the sensor needs to exchange a number of packets. A sequence of data and control packets to complete a task is called a Packet Flow (PF). Sensors can manage their power by defining priorities for tasks with the help an internal EC model. Moreover, minimization of power usage may be done by assigning optimum value to effective network parameters by developer with the help an external EC model. In this paper we focus on the internal EC modeling, and aim to model incoming tasks so that a sensor can prioritize them in a way that minimizes the EC.

The rest of the paper is structured as follows. Section II discusses related work on energy consumption models. Section III presents a discussion on Linear Modeling techniques. Our approach for modeling will be explained in section IV, and in section V we discuss an experimental result which shows the usefulness of the model. Finally, we summarize our work and outline future research directions in section VI.

## II. RELATED WORK

Network architectures such as the OSI and the Internet architectures are basically functional models organized as layers with the layer below provides services to the layer above, and eventually the application layer provides services to the end users. A network is often evaluated in terms of its quality of service parameters such as delay, throughput, jitter, availability, reliability, and even security. However, when it comes to the EC, one often encounters difficulty in the overall network evaluation and hence optimization as there hardly exists a model that takes the EC into account. Researchers fall back to the traditional network architecture and try to minimize selected components of a single layer with the hope that the overall EC of the network is reduced without regard for other components or layers. This is hardly an ideal situation where one does not know how a single component fits within the overall EC picture of an entire wireless sensor network.

Most current energy minimization approaches considered WSNs along the line of network layers: (1) the operating system, (2) the physical layer, (3) the MAC layer, (4) the network layer, (5) the application layer, and (6) the power harvesting layer. We propose a power consumption model for a deployed sensor in a WSN. This power consumption model provides a clear break down of the major constituents, which consume power of a sensor. The current model used by the sensor network community does not provide this level of insight, but instead mixes the sources of the power consumption together[7].

Other researchers focus exclusively on the cost of sending and receiving data to evaluate the EC of WSN[8].The energy required for transmitting or receiving a data bit is modeled as follows:

$$E_T(d) = \begin{cases} E_{T\_elec} + \varepsilon_{fs} \times d^2, d < d_0 \\ E_{T\_elec} + \varepsilon_{mp} \times d^4, d \geq d_0 \end{cases}$$

$$E_R = E_{R\text{-}elec}$$

Where, The electronics energy of transmitting and receiving, $E_{T\text{-}elec}$, $E_{R\text{-}elec}$, and $E_{elec}$, depends on factors such as the digital coding, modulation, filtering, and spreading of the signal, whereas the amplifier energy, $\varepsilon_{fs}$, and $\varepsilon_{mp}$, depends on the distance, $d$, to the receiver and the acceptable bit-error rate, if the distance is less than a threshold, the free space (fs) model is used; otherwise, the multipath (mp) model is used. Ref.[8] takes the characteristics of the power amplifier into account separately, but does not analyze the impact of the parameters on the transmission power and distance of communication. The power consumption model defined in [8],[9] as follows:

$$d \geq \left( \frac{E_{T\_elec} + E_{R\_elec}}{(1-2^{1-\alpha}) \times \varepsilon_{amp}} \right)^{1/\alpha}$$

Where $d$ is the distance between source *S* and destination *D*, then there is an intermediate node between *S* and *D* so that the retransmission will save energy. Other approaches evaluate the energy efficiency of a wireless sensor network by using the power consumption model mentioned in[8],[9]. For example,[10] uses the model to study energy efficient routing protocol; [11] uses the model to derive a cross design including physical, data link, and network layer.
Existing models consider transmitting and receiving activities, but parameters that quantify the tasks and the EC of these activities are not taken in to account. A sensor has to execute several tasks and consumes energy to run assigned tasks. The

EC will be recorded in term of tasks that a sensor executes. In this paper, the EC tasks are categorized in five main constituents: individual, local, global, sink, environment (figure1). The effective parameters of each constituent will be considered to evaluate the proposed sensors' EC model. Based on this model each node may be able to adjust its own parameters depending on its role within the network. Using the model for prediction, a sensor may devise its own policy and power management scheme for performing its tasks efficiently as part of minimization of the overall EC of the whole WSN application.

## III. ENERGY DRIVEN MODEL (EDM)

The current EC models are specified for the sensor network factors like radio[9], data[12], and hop[7], however, there are some other significant factors like number of packets a node creates, processes, transmits, receives, and senses etc. Moreover, wireless network characteristics are quite different from wire line systems. The wireless channel characteristics generally affect all the OSI layers. Manipulating a layer locally has direct influence on the EC of other layers in WSNs. Optimizing each layer individually to fix the problem leads to unsatisfactory results. It is argued in[13] that it is hard to achieve design goals like energy efficiency using the traditional layered approach. So the cross layer was created to enhance the performance of the system by jointly optimizing multiple protocol layers[6]. It is argued that Cross-Layer Designs with tight coupling between the layers become hard to review and redesign. Changing one subsystem implies changes in other parts, as everything is interconnected. Moreover, Cross-Layer Designs without solid architectural guidelines inevitably reduced flexibility, interoperability and maintainability. In addition, systems may become unpredictable. It is hard to foresee the impact of modifications.

In this paper, we create a new modular view involving energy consuming constituents (figure 1). We propose an approach for modeling the overall EC in terms of effective parameters and energy consuming constituents.

We consider five energy consuming constituents (figure 1) based on their tasks as shown in figure 2.

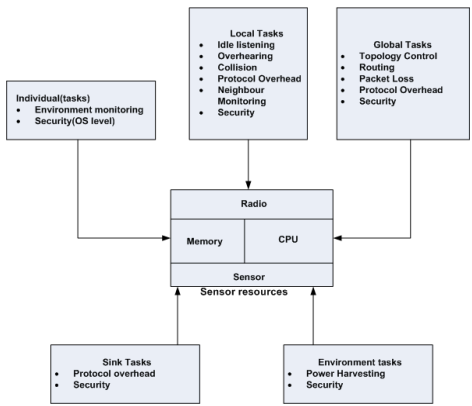

Figure 2. System design in term of constituents' tasks.

The Individual constituent defines all the essential and basic operations or tasks for the sensor to just exist i.e. monitoring environment events as a key task of a sensor, executing OS and providing security in OS level. The Local constituent deals with initiating and maintaining all communications with a node's immediate neighbors i.e. monitoring neighbors and providing a secure communication with neighbors at the local level. In addition, it may include the power usage of overhearing, idle listening and collision if they happen. The Global constituent is concerned with the maintenance of the whole network, the selection of a suitable topology and an energy efficient routing strategy based on the application's objective. This may include energy wastage from packet retransmissions due to congestion and packet errors. The global constituent is defined as a function of the EC for topology management, packet routing, packet loss, and protocol overheads. Sink constituent assumes the roles of manager, controller or leaders in WSNs. The sink tasks include directing, balancing, and minimizing the EC of the whole network, and collecting the generated data by the network's nodes. The Environment tasks consist on deploying the energy-harvesting operations in the case where nodes have the capability of extracting energy from environment. Execution of these tasks requires sensor resources, CPU, Memory, Radio, and Sensor.

Having knowledge of costly functions can guide a sensor to run tasks based on its residual power and the importance of tasks. Establish a balance between the EC of constituents can also guide a sensor to minimize the EC. So from sensor view point, the challenge will be selecting and executing efficiently significant tasks in the optimum order to minimize its EC. In addition, moving tasks from a constituent with high level EC to a low level EC constituent can minimize the EC e.g. data aggregation that reduce global tasks and increase individual tasks. Moreover, we may split a high EC task to low level tasks that are suitable for low EC constituents. So sensors can act intelligently to manage task execution in the efficient way based on the EC model.

In the following, we show the relation between significant parameters of the task basis constituents and the overall EC. Generally when a sensor runs a typical task, the energy will be consumed by CPU, Memory, Radio and Sensor units we show it as:

$$e_t = e_{cpu} + e_{mem} + e_R + e_{sens} \qquad (1)$$

For each task, a sensor runs the basic operations. We assume a sensor as a server that should execute incoming tasks. Since the common and primary resources in all type of sensors are CPU, Memory, Radio and Sensor, we assume the EC of sensors according to[14]. In[14], they purposed an approach to model the EC From hardware perspective:

$$e_{cpu} = p_{cpu} b_{cpu} \qquad (2)$$
$$e_{mem} = p_{mem} b_{mem} \qquad (3)$$
$$e_{Radio} = e_{Tx} + e_{Rx} \qquad (4)$$
$$e_{Rx} = p_{Rx} b_{Rx} \qquad (5)$$
$$e_{Tx} = p_{Tx} b_{Tx} \qquad (6)$$
$$e_{sens} = p_{sens} b_{sens} \qquad (7)$$

Where $b_{cpu}$, $b_{mem}$, $b_{Rx}$, $b_{Tx}$, $b_{sens}$ shows number of packets processed in CPU, stored in Memory, Received, Transmit by Radio, and sensed, respectively.

Every task that sensor should do in its lifetime is assigned to a constituent. Obviously the overall EC of a typical sensor can be calculated by the power usage of doing the individual, local, global, environment, and sink tasks (figure 1):

$$E_{Overall} = E_{Individual} + E_{local} + E_{global} + E_{environment} + E_{Snk} \quad (8)$$

Since each constituent includes a number of tasks and tasks include a PF (*b*), the EC of a sensor is as follows:

$$E_{Overall} = \underbrace{(\lambda_1 p_{cpu} + \lambda_2 p_{mem} + \lambda_3 p_{Rx} + \lambda_4 p_{Tx} + \lambda_5 p_{sens})}_{\alpha_1} b_{Individual} +$$
$$\underbrace{(\lambda_6 p_{cpu} + \lambda_7 p_{mem} + \lambda_8 p_{Rx} + \lambda_9 p_{Tx} + \lambda_{10} p_{sens})}_{\alpha_2} b_{local} +$$
$$\underbrace{(\lambda_{11} p_{cpu} + \lambda_{12} p_{mem} + \lambda_{13} p_{Rx} + \lambda_{14} p_{Tx} + \lambda_{15} p_{sens})}_{\alpha_3} b_{global} +$$
$$\underbrace{(\lambda_{16} p_{cpu} + \lambda_{17} p_{mem} + \lambda_{18} p_{Rx} + \lambda_{19} p_{Tx} + \lambda_{20} p_{sens})}_{\alpha_4} b_{environment} +$$
$$\underbrace{(\lambda_{21} p_{cpu} + \lambda_{22} p_{mem} + \lambda_{23} p_{Rx} + \lambda_{24} p_{Tx} + \lambda_{25} p_{sens})}_{\alpha_5} b_{snk} \quad (9)$$

Therefore:
$$E_{Overall} = \alpha_1 b_{Individual} + \alpha_2 b_{local} + \alpha_3 b_{global} + \alpha_4 b_{environment} + \alpha_5 b_{snk} \quad (10)$$

TABLE 1. INDIVIDUAL EFFECTIVE PARAMETERS ON ENERGY CONSUMPTION

| | Individual Parameters | | |
|---|---|---|---|
| index | Parameter | Description | Boundary |
| 1 | $r_{sense}$ | Sensing radius points to the covered area of the sensor, this will have different meaning in different applications e.g. a temperature application and a radar application. | $r_{sense} > 0$ |
| 2 | $g_{sense}$ | Sensing delay | $g_{sense} \geq 0$ |
| 3 | $b_{sense}$ | Number of packet created by sensor itself that includes environment's data. | $b_{sense} \geq 0$ |
| 4 | $b_{store}$ | Numbers of packets are stored in the memory. | $b_{store} \geq 0$ |
| 5 | $b_{Os}$ | Number of Os instruction | $b_{OS} \geq 0$ |
| 6 | $b_{sec}$ | Security in Individual level | $b_{sec} \geq 0$ |

In the following sections, we explain each constituent in term of effective parameters on sensor's EC model:

*A. Individual*

$b_{Individual}$ consists of PF of individual tasks i.e. sensing task, executing OS and installed applications and also providing security for a sensor individually[15-16]. Therefore PF in the Individual constituent is:
$$b_{Individual} = b_{sens} + b_{OS} + b_{sec} \quad (11)$$
Number of sensed and produced packets by a sensor depends on the covered area by a sensor, $r_{sens}$, and sensing delay, $g_{sens}$, therefore:
$$b_{sens} = P[Sense|r_{sens}, g_{sens}]b_{Individual} \quad (12)$$
According to eq. 11 and eq. 12:
$$b_{Individual} = \frac{b_{OS} + b_{sec}}{1 - P[Sense|r_{sens}, g_{sens}]} \quad (13)$$

*B. local*

$b_{local}$ includes PF for neighbor monitoring to gather information of neighbor's available resources such as the residual energy and the memory space, the security management to prevent malicious nodes from destroying the connectivity of the network and tampering with the data, idle listening packets, overhearing packets, retransmission packets due to collision and the tasks to prevent them. Therefore the local constituent's packet flows can be as:

$$b_{local} = b_{coll} + b_{idle} + b_{ohear} + b_{sec} + b_{mon} + b_{ohead} \quad (14)$$
Where:
$$b_{coll} = P[coll|n, g_{Tx}, net_{dens}]b_{local} \quad (15)$$
$$b_{ohear} = P[ohear|n, net_{dens}, r_{Tx}]b_{local} \quad (16)$$
$$b_{idle} = P[idle|n]b_{local} \quad (17)$$

Where *n* is the number of neighbors, $net_{dens}$, is the total number of nodes in the network, $g_{Tx}$ is the transmission delay and , $r_{Tx}$, is the transmission radius. Therefore according to eq.14, 15, 16, and 17:

$$b_{local} = \frac{b_{sec} + b_{mon} + b_{ohead}}{1 - (P[coll|n, g_{Tx}, net_{dens}] + P[ohear|n, net_{dens}, r_{Tx}] + P[idle|n])} \quad (18)$$

*C. global*

The Global constituent consists of a number of tasks: topology control, routing, retransmission due to the packet loss, and performing tasks to prevent the pack loss.
$$b_{global} = b_{pktls} + b_{sec} + b_{topo} + b_{rout} + b_{ohead} \quad (19)$$
The possibility of the packet loss in the network depends on the effective parameters like D, distance between node and destination, and $net_{dens}$, number of nodes in the network.
$$b_{pktl} = P[pktls|D, net_{dens}]b_{global} \quad (20)$$
According to 19 and 20:
$$b_{global} = \frac{b_{sec} + b_{topo} + b_{rout} + b_{ohead}}{1 - P[pktls|D, net_{dens}]} \quad (21)$$

*D. environment*

The Environment constituent includes providing the security and the power harvesting management if a node has ability to harvest energy from the environment:
$$b_{environment} = b_{sec} + b_{ph} \quad (22)$$

*E. Sink*

The Sink constituent includes providing the security for the

TABLE 2. LOCAL EFFECTIVE PARAMETERS ON ENERGY CONSUMPTION

| | Local Parameters | | |
|---|---|---|---|
| index | Parameter | Description | Boundary |
| 1 | n | Number of neighbors | $n \geq 1$ |
| 2 | $e_i(idle)$ | Idle power consumption | |
| 3 | $d_{ij}$ | Distance to the neighbor | $0 < d_{ij} \leq r_{Tx}$ |
| 4 | $b_{mon}$ | Packet overhead for monitoring depends on the application and its topology. | $b_{mon} \geq 0$ |
| 5 | $r_{Tx}$ | Transmission Radius | $r_{Tx} \geq 0$ |
| 6 | $b_{sec}$ | Local Security packet overhead depends on application. | $b_{sec} \geq 0$ |
| 7 | $b_{local}$ | Packet overhead to avoid collision problem policy. | $b_{local} \geq 0$ |
| 8 | $b_{reTx}$ | Number of retransmission packets depends on probability of collision and number of neighbors | $b_{reTx} \geq 0$ |
| 9 | $b_{sec}$ | PF security in local level | $b_{sec} \geq 0$ |

TABLE 4. ENVIRONMENT EFFECTIVE PARAMETERS ON ENERGY CONSUMPTION

| | Environment Parameters | | |
|---|---|---|---|
| index | Parameter | Description | Boundary |
| 1 | $H_i$ | Harvested energy (Wat) | $H_i \geq 0$ |
| 2 | $b_{ph}$ | Overhead produced due to harvesting power. | $b_{ph} \geq 0$ |

TABLE 5. SINK EFFECTIVE PARAMETERS ON ENERGY CONSUMPTION

| | Sink Parameters | | |
|---|---|---|---|
| index | Parameter | Description | Boundary |
| 1 | $b_{ohead}$ | Network management policy | $b_{ohead} \geq 0$ |
| 2 | $b_{sec}$ | PF security in sink level | $b_{sec} \geq 0$ |

sink communication and performing sink directions if it is applicable in the application:

$$b_{snk} = b_{sec} + b_{ohead} \qquad (23)$$

Tables 1-5 show the constituents parameters. We use linear modeling and the regression to establish the relation between above parameters and the EC. So we use usual regression method to calculate $\alpha_1, \alpha_2, \alpha_3, \alpha_4, \alpha_5$ of the equation 10. In the following, we explain our method to learn the EC model from the experiments.

## IV. ESTIMATING MODEL COEFFICIENTS USING REGRESSION

Taking multiple observations of the observable quantities allows estimating the model parameters using learning techniques such as linear regression. We use the linear regression with ordinary least square estimation. To generate a sufficient number of observations resulting in linearly independent equations and spanning a large range of packet flows, we load sensor using constituents' packet flows. The model uses the parameters at run time to control the constituents' power usage.

### A. Least-Square (LS) approximation

We assume constituent's packet flows in the application and do experiments. To have more precise model, number of experiment (M) should be very bigger than number of Constituents (N), (M>>N) as[17-19]:

$$\overbrace{\begin{bmatrix} E_1 \\ E_2 \\ \vdots \\ E_M \end{bmatrix}}^{E} = \overbrace{\begin{bmatrix} b_{Ind}^{(1)} & b_{local}^{(1)} & b_{global}^{(1)} & b_{snk}^{(1)} & b_{env}^{(1)} \\ b_{Ind}^{(2)} & b_{local}^{(2)} & b_{global}^{(2)} & b_{snk}^{(2)} & b_{env}^{(2)} \\ & & \vdots & & \\ b_{Ind}^{(M)} & b_{local}^{(M)} & b_{global}^{(M)} & b_{snk}^{(M)} & b_{env}^{(M)} \end{bmatrix}}^{b} \overbrace{\begin{bmatrix} \alpha_1 \\ \alpha_2 \\ \vdots \\ \alpha_5 \end{bmatrix}}^{A}$$

$$\Rightarrow A = ? \qquad (24)$$

Where $E$ matrix is the EC in different observation and b matrix is the number of packets for constituents' tasks. For example, the number of packets for the Individual, local, global, sink, environment tasks in $n^{th}$ observation it is shown by $b_{Ind}^{(n)}$, $b_{local}^{(n)}$, $b_{global}^{(n)}$, $b_{snk}^{(n)}$, $b_{env}^{(n)}$, respectively. $A$ matrix is a coefficient matrix that should be calculated for the model. The Least Square approximation says if $A = (b^T b)^{-1} b^T E$ then E is calculated for new values of the parameters by inner product of A and P: $E = Ab$.

### B. Refine the model

The problem with the model however is the linearity does not necessarily hold across the constituents, since the constituents do not have homogeneous packet flows, or in other words, number of packets in order to complete a task is not similar. The PF of a task is a significant concept, network protocols directly effect on the number of needed packets (PF) to complete a task. A senor determines a PF for each task based on the average number of sent and received control and data packets in the first execution of each task. In addition, each sensor has a different model due to different constituents' tasks, for example, sensors near to sinks have more global tasks than those far from sinks.

Values are collected in time periods (slices), Δt, and the model should be repeated in a number of time slices to determine the unknown parameters for the constituents because constituents' tasks and the constituents parameters' values change on time, for example, a sensor can be head cluster or just acts as an accelerator to monitor the environment, or in other words, its tasks can be changed several times in its life time. Repeating modeling helps to have knowledge of the cost of constituents' tasks, and then a sensor can decide based on its power model which tasks should be run to have a longer lifetime.

## V. EXPERIMENT

### A. Experiment setting

We have simulated a WSN application to track the EC of constituents. The application collects information about events that occur. Sensors detect an event in their covered area and create a packet and send it to the nearest sink. Sinks are located as a group in specific location. Generally we assume three phases in our WSN application simulator. In the Initialization phase, a sensor executes its own software, creates connection with immediate nodes as a neighbor and collects information about the neighbor's resources. Then in the Collecting phase, the sensor uses neighbor's information to relay data. Moreover, in the data collection phase, the sensor collects information from the environment and creates data and sends them, and also it has to process and relay incoming packets. It performs these tasks if it has enough power otherwise it ignores them. In the Maintenance phase, the sensor monitors its neighbors to update their situation, and it has to perform extra global tasks such as reorganizing topology and reconfiguring routing tables when it is necessary. These phases may be repeated by a sensor a number of times during the network life time.

Table 6 shows how we assign PFs to constituents in the simulator. In our application, sensors have connection with all immediate nodes and they always select neighbors based on their residual energy. The sink does not have any roles in the application, and sensors do not harvest energy hence we do not consider the Sink and the Environment constituents in the EC modeling. According to table 1, the sensor distinguishes Packet Flows of different constituents. In the next section, the method of learning $\alpha_1, \alpha_2, \alpha_3, \alpha_4$ and $\alpha_5$ of eq. 10 from experiments is explained.

### B. Results

In this section, we investigate various packet flows and the energy consumptions of each constituent in all phases of our WSN application through simulation experiments. The first phase covers time slices before a sensor starts sensing, monitoring and relaying data packets. In this experiment, the first phase consists of three time slices. A sensor spends power to start up (Individual), transport control packets, and initialize connections with neighbors (Local). It also sends control packets to set routing tables (global). In the second phase, the sensor starts capturing events and creates data packets and sends them to a sink (Individual tasks). As local tasks, it monitors neighbors' resources by sending request packets to its neighbors. Moreover, the sensor is responsible for relaying incoming data packets to their destination by looking at its routing table and choosing a suitable neighbor or path. The third phase starts when the network needs to recover from a disconnected path. In this phase, the sensor performs the second phase tasks; also it has to do extra global tasks to maintain the network. These tasks involve capturing information about paths and updating routing tables.

TABLE 6. PACKET FLOW OF DIFFERENET CONSTITUENTS

| Packet Flow | Constituent |
|---|---|
| Sensed Packets | Individual |
| Packets carrying neighbor and node's current information | Local |
| Scheduling Packets to avoid Collision | Local |
| Packets carrying topology information | Global |
| Packets carrying routing information | Global |
| Received data packets | Global |

Figures 3a and 3b show the PF and the EC of each constituent, respectively, of a typical sensor in different phases (initialization, data collection and maintenance) of simulation experiments. In each time slice the PF of constituents (based on eq.10 to eq.21) and the EC (based on current sensor's power level) are recorded. As can be seen in phase 3, an increase in the Global constituent activities resulted in a drastic increase in the overall EC. Figure 3a and 3b in phase 1 and 2 show that variations of the Individual and Local constituents' tasks do not have considerable effects on the overall EC. It can be clearly seen from Figure 3a and 3b that the increase of global tasks results in a peak in the EC as shown in time slices of phase 3. The Global tasks (from the Global constituent) are thus very costly in terms of the EC in the sensor life time and directly affect the overall EC of the WSN application.

In the simulation, we only considered a simple routing protocol based on the residual energy of neighbors. We did not assume packet loss, overhead of a topology control, and security protocols, however, the Global constituent still has a massive influence on the node's energy usage. If more complex protocols are deployed that entail heavy control packet flows to global tasks, the global constituent would become the dominant constituent in term of the overall energy usage of the node and hence the energy consumption of the overall WSN application.

We model the overall EC of a sensor using linear regression as

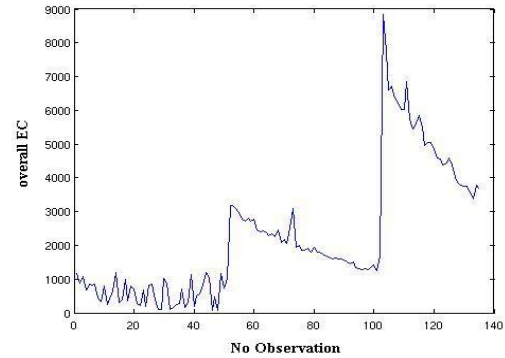

(a)

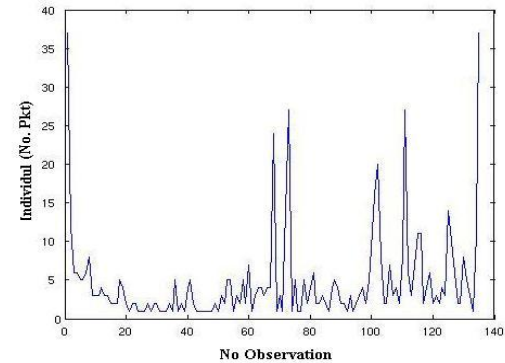

(b)

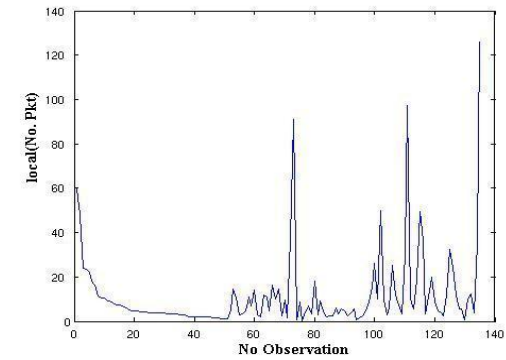

(c)

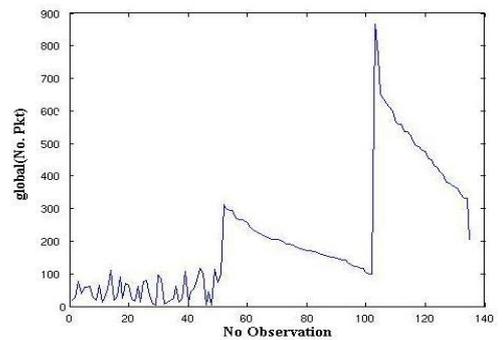

(d)

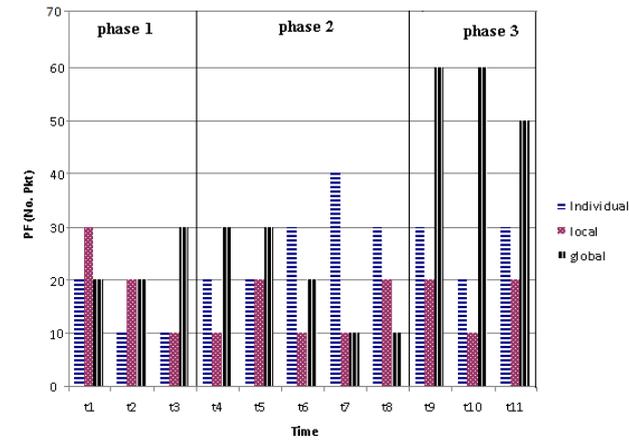

(a)

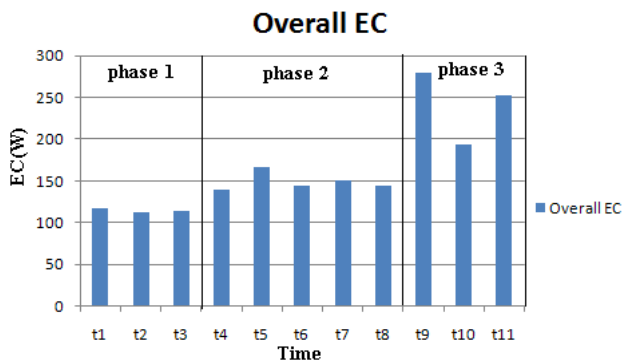

(b)

Figure 3. (a) constituent's packet flows in different time slice (b) overall EC in different time slices.

Figure 4. Number of packets of Individual, local, and global and Overall EC in different observation

follows:
$$E_{overall} = \alpha_1 b_{Ind} + \alpha_2 b_{local} + \alpha_3 b_{Global} \quad (25)$$
$\alpha_1$, $\alpha_2$, and $\alpha_3$ are learned from many experiments with random values for configurable parameters in Tables 1 to 3. Figure 4a shows the overall EC of a sensor in different observations. Figures 4b to 4d show the PF of the Individual, the Local and the Global constituents in different experiments. We recorded the overall EC and constituents' PFs in order to learn the model's coefficients. We compared the variations of the EC in figure 4a and variations of constituents' PF in Figure 4b to 4d. The results indicate that the Global constituent is clearly the most dominant constituent of the overall EC. It can clearly been seen from figure 4a and 4d, they change similar to each other. To test the accuracy of our EC model, we use it to predict the EC of a typical sensor in a number of simulation experiments with random values within the predefined range of effective parameters of the Individual, the Local, and the Global constituents. We then run experiments on the simulator and capture the EC to determine the prediction error (Figure 5a). Figure 5b shows the prediction accuracy of our application by comparing the actual EC and its predicted value. We found that the average error between the observed EC and the predicted values is about 13%. The errors are expected partly from the model inaccuracy and partly from the linearity assumption. As can be seen from figure 5b, there are some spikes in the prediction errors. These spikes generally happen in the high values of the EC which probably caused by large differences of the EC of the Global constituent in comparison with other constituents. Figure 5c shows that larger values of the Global constituent imply higher error in the EC prediction. It is expected that the obtained model cannot predict the EC of a sensor perfectly, but it can clearly reveal the relationship among the energy consuming constituents of a sensor.

## VI. DISCUSSION AND FUTURE WORK

Modeling the overall energy consumption as a linear combination of its EC constituent is clearly a first order approximation of the energy consumption pattern of a sensor; however, clearly the model can identify and separate major constituents. With this knowledge, a sensor may be attributed its own energy management policy. Moreover, by finding an appropriate mathematical relation between constituent's parameters and resources' utilization (CPU, Memory, Radio, and Sensor), it is possible to turn the problem of finding the best values of configuration parameters into an optimization problem:

$$Min\ E_{Overall} = E_{Individual} + E_{Local} + E_{Global} + E_{environment} + E_{Snk} \quad (26)$$
$$subject\ to:$$
$$1: E_{local} > 0$$
$$2: E_{global} > 0$$
$$3: E_{individual} + E_{local} + E_{global} + E_{snk} < E_{battery}$$

The plan for future is to model the constituents with respect the most effective parameters and optimize each constituent by considering optimum values for parameters. In addition we will plan to model the EC of whole network to achieve an optimum solution for task assignment. We expect that the result of optimization of constituents affects on the Overall EC and the number of tasks of the constituents will be minimized

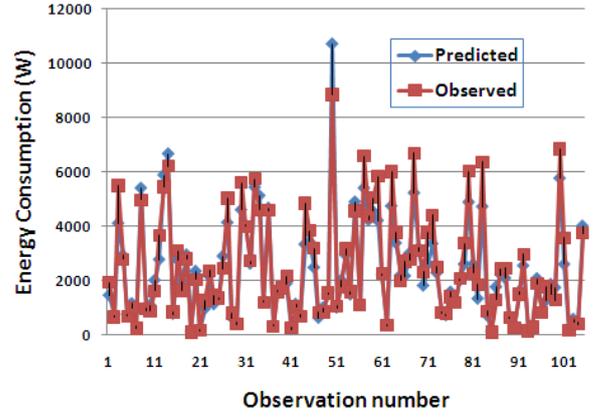

(a)

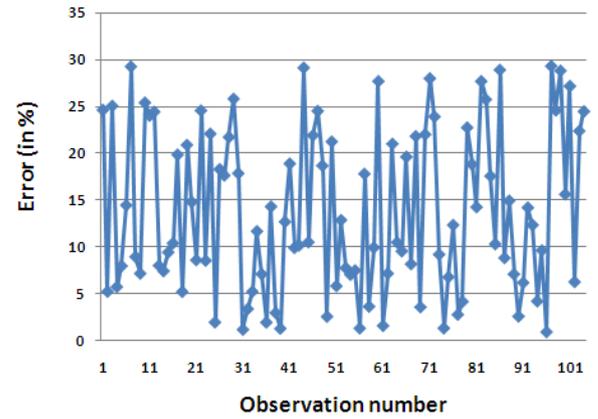

(b)

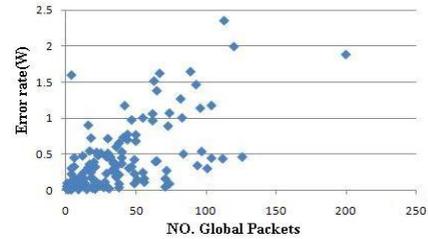

(c)

Figure 5. (a)Model predicted and observed values of EC for several random runs. (b)The error range of the model(c) comparing the variation of the Error value with global packet flows.

and the procedure of the task selection and also the task assignment can be controllable.

## VII. CONCLUSION

The motivation behind this work is the need to minimize the overall EC of sensors. We introduce five energy consuming constituents: individual, local, global, environment and sink. Each constituent consists of a set of tasks based on the wireless sensor application characteristics. Our model helps identify essential EC constituents and their contribution to the overall energy consumption of a sensor. This in turn helps the sensor to spend its energy wisely. The sensor extracts/profiles constituent's power usages and then applies the regression to establish a relationship between constituents' tasks and the overall EC. The model can then be utilized by the sensor to prioritize the constituents' tasks in term of the EC level and importance in order to make appropriate decision. So the sensor can use the power in an effective way and remain alive longer. Using the same model for extracting its power consumption profile, a sensor equipped with an intelligent algorithm can even act appropriately to conserve its energy in power shortage situations. We call these sensors "thrifty sensor" and the idea of thrifty sensors is worth exploring in the future.

We expect the system design and modeling procedure is valid for various applications. It means the EC of WSN applications can be modeled with respect constituents. It is obvious that the coefficients of the modeling changes from one sensor to another and from one application to another application.